\documentclass[12pt,preprint]{aastex}

\begin{document}
\title{{\it GALEX} and Optical Light Curves of WX LMi, SDSSJ103100.5+202832.2
 and SDSSJ121209.31+013627.7\footnote{Based on observations made with the NASA 
Galaxy Evolution Explorer.
{\it GALEX} is operated for NASA by the California Institute of Technology under
 NASA contract NAS5-98034.}}

\author{Albert P. Linnell\altaffilmark{2},
Paula Szkody\altaffilmark{2}, 
Richard M. Plotkin\altaffilmark{2},
Jon Holtzman\altaffilmark{3},
Mark Seibert\altaffilmark{4},
Thomas E. Harrison\altaffilmark{3},
Steve B. Howell\altaffilmark{5}}
\altaffiltext{2}{Department of Astronomy, University of Washington,
Box 351580,
Seattle, WA 98195, szkody@astro.washington.edu,plotkin@astro.washington.edu,
linnell@astro.washington.edu}
\altaffiltext{3}{Department of Astronomy, New Mexico State University, Box 30001,
MSC 4500, Las Cruces, NM 88003, tharriso@nmsu.edu,holtz@nmsu.edu}
\altaffiltext{4}{Observatories Carnegie Institute of Washington, 813 Santa Barbara
St., Pasadena CA 91101, mseibert@ociw.edu}
\altaffiltext{5}{National Optical Astronomy Observatories,
950 N. Cherry Avenue, Tucson, AZ 85726, howell@noao.edu}

\begin{abstract}

{\it GALEX} near ultraviolet (NUV) and far-ultraviolet (FUV) light curves
of three extremely low accretion rate polars show distinct modulations in
their UV light curves. While these three systems have a range of magnetic
fields from 13 to 70 MG, and of 
late type secondaries (including a likely brown dwarf in
SDSSJ121209.31+013627.7), the accretion rates are similar, and the UV
observations imply
some mechanism is operating to create enhanced emission zones on the white dwarf.
The UV variations
match in phase to the two magnetic poles viewed in the optical in WX LMi
and to the single poles evident in
the optical in SDSSJ1212109.31+013627.7 and SDSSJ103100.55+202832.2.
Simple spot models of the UV light curves show that if hot spots are
responsible for the UV variations, the temperatures are on the order
of 10,000-14,000K. For the single pole systems, the size of the FUV
spot must be smaller than the NUV and in all cases, the geometry is
likely more complicated than a simple circular spot.

\end{abstract}


\section{Introduction}

The cataclysmic variables with very low mass transfer rates 
(10$^{-13}$-10$^{-14}$ M$_{\odot}$ yr$^{-1}$) and highly
magnetic ($>$10MG) white dwarfs have been termed LARPS 
(Low Accretion Rate Polars;
Schwope et al. 2002). Since the low mass transfer means the accretion 
luminosity is low,  the individual stellar components can be
viewed. Thus, these systems
 provide a unique comparison to widely separated binaries for the study of how
angular momentum losses and heating of the white dwarf affect the binary
evolutionary scenario.
While normal polars can reach mass accretion rates close to those of LARPS 
when they undergo long periods of low mass transfer, the main
difference between LARPS and normal polars in low states appears to be
the temperature of the white dwarf. Ultraviolet observations of normal
polars have shown temperatures of 11-14,000K for their white dwarfs
(Araujo-Betancor et al. 1995). In contrast, the white dwarfs in the
LARPS
are less than 10,000K (Schmidt et al. 2005, 2007, hereafter S05, S07; 
 Vogel et al. 2007, hereafter 
V07;
Schwope et al. 2009, hereafter Sw09). These low temperatures imply that compressional heating
has not taken place and so these objects have been suggested to
 be pre-Polars (S05)
 or PREPs (Sw09). Figure 5 of Sw09 shows a nice plot of the temperatures of
9 PREPs compared to normal polars and accreting non-magnetic white dwarfs that
illustrates this difference. However, the terminology and behavior is not
always so cleanly separated. The system SDSS121209.31+013627.7 (Schmidt et al. 2005b;
Burleigh et al. 2006; hereafter B06) also has a low accretion rate and a low white
dwarf temperature but has a brown dwarf secondary. Its evolutionary path has been
suggested to be either a polar in a low state or a PREP. In addition, the recent
observation of enhanced activity levels in the LARP SDSSJ204827.9+005008.9
(Honeycutt et al. 2010) suggest that it may not be a LARP/PREP or that even PREPs
can have different levels of activity. 

A secondary criteria for a LARP/PREP is the secondary underfilling its Roche lobe
so that Roche lobe overflow does not occur (S05; S07; Sw09). This is generally
determined from the system having an spectral type too late to fill the lobe if it has
a normal size for its type, as well as the absence of narrow components in
the emission lines that can be traced to a stream.  However, the Honeycutt et al.
(2010) study found evidence for a stream in SDSSJ204827.9+005008.9, although they
could not determine if the system had entered an increased state of mass transfer
at the time.
While the low accretion rates and sizes of the secondaries
imply no Roche lobe overflow, there could likely still be some accretion from
the stellar wind of the secondary that is funnelled by the high field of
the white dwarf (Li et al. 2004). The work of Webbink \& Wickramasinghe (2005)
using considerations of energy, magnetic fields, separations and lifetimes supports
a wind model for the LARPS.  In this study, we are not so concerned
with the correct classification of our objects as LARPS, PREPs or normal polars in low states, or how the accretion might occur,
but rather with the effects of
low levels of accretion on the white dwarf.

These effects are visible through time-resolved
UV observations.  Accretion spots with temperatures of 30,000-70,000K are easily
 visible on normal polars even during their sporadic low states of accretion
with $\dot{M}\sim$10$^{-12}$M$_{\odot}$ yr$^{-1}$ (Araujo-Betancor et al.
2005; G\"ansicke et al. 2006) as large amplitude modulations of the UV
light. However, it was somewhat surprising to
find that similar modulations were apparent for EF Eri (Szkody et al. 2006)
which had been in a low state for seven years with $\dot{M}$ of 10$^{-13}$ M$_{\odot}$ yr$^{-1}$ and the LARP MQ Dra with an even lower 
$\dot{M}$ of 10$^{-14}$ M$_{\odot}$ yr$^{-1}$ (Szkody
et al. 2008). While the UV light curves could be approximately matched
with either hot spots or cyclotron components (Campbell et al. 2008), each
interpretation
has its problems: the hot spots require different sizes and geometries while
the cyclotron origin requires much higher magnetic fields in the UV than
are apparent in the optical.
In order to further study the effects of this low level accretion, we
used {\it GALEX} to obtain UV observations of two additional
LARPS (WX LMi and SDSSJ103100.55+202832.2) and the system  
SDSSSJ121209.31+013627.7. 
For convenience, we will refer to the SDSS objects as SDSS1031 and
SDSS1212. The known parameters for the three systems are listed in
Table 1. While all three of these objects have comparable low mass
accretion rates, they present differences in magnetic field strength,
white dwarf temperature, spectral type of secondary and orbital period.

\section{Observations}
The observational data include ground-based APO $B,V,R,I$ light
 curves and FUV and NUV {\it GALEX} light curves.
The {\it GALEX} observations took place on 2008 Feb 16, 23 and April 5
(Table 2 provides a summary of the times).  Images in both the near 
ultraviolet (NUV)
detector (1750-2800\AA) and the far ultraviolet (FUV) detector (1350-1750\AA)
(Martin et al. 2005) were obtained for each source. The time-tag data
were calibrated in 240s intervals (120s for the NUV of SDSSJ1212) 
and phased according to the spectroscopic
ephemeris given
in V07 for WX LMi, and to the photometric ephemeris given 
in B06 for SDSS1212. The phasing for SDSS1212 is accurate to $\pm$0.14 phase
at the time of the {\it GALEX} obseravtaions. For SDSS1031, the ephemeris 
is not known with enough accuracy to phase to prior data so arbitrary 
phasing was used with the period of
S07. The phased datapoints were then
binned into 10, 15 and 20 phase bins to optimize the S/N vs time resolution.
The  magnitudes for each binning were then measured using a 9 pixel radius
aperture with the IRAF\footnote{IRAF (Image
 Reduction and Analysis
Facility) is distributed by the National Optical Astronomy Observatories, which
are operated by AURA,
Inc., under cooperative agreement with the National Science Foundation.} 
routine
{\it qphot}, using an average sky over the entire phase interval. Flux
conversions for the modeling were done using the values given in the 
{\it GALEX} online documentation\footnote{See http://galexgi.gsfc.nasa.g
ov/tools/index.html} (FUV m$_{0}$=18.82=1.40$\times$10$^{-15}$ergs cm$^{-2}$ s$
^{-1}$ \AA$^{-1}$
and NUV m$_{0}$=20.08 =2.06$\times$10$^{-16}$ ergs cm$^{-2}$ s$^{-1}$ \AA$^{-1}
$). 

Optical photometry in 2008 April and May (Table 2) was obtained for WX LMi and
SDSSJ1031 using the NMSU 1m telescope at Apache Point Observatory.
Differential light curves in $BVRI$ filters were made from nearby
comparison stars on the same CCD frames. For WX LMi, one of the comparison
stars used was the calibrated star A from Schwarz et al. (2001) so that the 
differential magnitudes could be transformed to actual $BVRI$ values.

\section{The Analysis Program}

As in our previous spot models of EF Eri and MQ Dra, the BINSYN program
suite (Linnell \&
Hubeny 1996) was used to calculate synthetic light curves that could
fit the observations. Recent papers (Linnell et al. 2007; 2008) illustrate
details of its application to cataclysmic variables, and Hoard et al. (2005)
describe its use in simulating hot spots on the magnetic white dwarf YY Dra. 
While a hot spot may be multi-temperature and have a complicated shape as
well as radiation influenced by the strong magnetic field, we
attempted to minimize the free parameters by using a first approximation
to the spot parameters
 with a circular, isothermal hot spot (with a white dwarf spectral energy distribution). As our
intent is to determine if a hot spot alone can account for the FUV and NUV
light curves, we ignore cyclotron effects. The harmonics will affect the
optical light curves to various degrees (depending on where they appear in
the various filters) and can add to or dominate the observed amplitude
of variation from the spot.

 In general, if the objective of the analysis is only to fit the light
variations (the absolute fluxes are unknown), the light curves 
are normalized to a light maxima
of 1.0,  and then BINSYN can be used 
to calculate monochromatic light
curves. In this mode, the program represents the stellar components
as a sum of black body contributions from local
photospheric values of $T_{\rm eff}$ determined by adopted values
of bolometric albedos and gravity darkening
exponents. No synthetic spectra are needed. 

However, the 
WX LMi light curves required special considerations, since those light
curves are absolute flux measurements.
 Their simulation,
via synthetic photometry from synthetic spectra,
produced calculated absolute flux values outside the Earth's atmosphere, using
transmission profiles of the filters and the quantum efficiency of the CCD, and
the distance of the system. 
In previous papers the synthetic spectra input to BINSYN were calculated using TLUSTY (Hubeny 1988)
and SYNSPEC (Hubeny et al. 1994). However, the Hubeny 
TLUSTY WD models do not extend to cooler than 8000K. 
Fortunately, cool WD models have been calculated by Bergeron et al. (1991; 1995).
Dr. Bergeron
kindly provided a set of 40 WD synthetic spectra with log~$g$=8.0 and $T_{\rm eff}$ ranging from 17,000K to 1500K.
Each synthetic spectrum includes Eddington flux values ($\rm erg~sec^{-1}~cm^{-2}~hz^{-1}$)
at 1788 unequally-spaced wavelengths between 43\AA~and 
95,000\AA.
Synthetic spectra similar to these have been used in the synthetic WD photometry of Holberg \& Bergeron (2006).
We produced corresponding synthetic spectra in wavelength units ($\rm erg~sec^{-1}~cm^{-2}~\AA^{-1}$).

We input WD synthetic spectra with $T_{\rm eff}$ values of 6000K, 6500K, 7000K, 7500K, and 8000K to BINSYN;
the program then used 
an adopted WD $T_{\rm eff}$, within the above range, and interpolated among the input
spectra to produce a synthetic WD spectrum of the assigned $T_{\rm eff}$. As representation of the hot spots on the WD
(the impact regions of the incoming mass from the secondary) requires additional synthetic WD spectra, we
added input WD synthetic spectra with $T_{\rm eff}$ values of 10,000K, 11,000K, and 12,000K. 

Synthetic spectra to represent the secondary star present a separate problem. If the secondary fills its
Roche lobe, $T_{\rm eff}$ and log~$g$ vary over the photosphere as determined by the values of gravity
darkening and bolometric albedo (temperature variation due to irradiation by the WD is negligible).
As BINSYN needs synthetic spectra for cool
photospheres over a range of $T_{\rm eff}$ and log~$g$, we downloaded a database 
of 
MARCS~\footnote{http://marcs.astro.uu.se}
synthetic spectra for
this purpose. Each MARCS synthetic spectrum tabulates physical flux values ($\rm erg~sec^{-1}~cm^{-2}~\AA^{-1}$)
for 101,125 wavelengths unequally
spaced between 900\AA~and 200,007\AA\ and we then prepared corresponding spectra in Eddington flux units.
To permit accurate interpolation to the local photospheric values of
$T_{\rm eff}$ and log~$g$ on the secondary star we provided BINSYN with 24 MARCS spectra for $T_{\rm eff}$
values of 2500K, 2700K, 2900K, 3100K, 3300K, and 3400K, and for each of these, log~$g$ values of 3.0, 4.0,
4.5, and 5.0. The single secondary component point nearest the L1 point has an associated photospheric segment 
whose flux contribution is negligible. We include an input black body spectrum of 1500K which, together with the
other input spectra for the secondary star, brackets this point. MARCS spectra span all other
grid segments covering the secondary star. 
The BINSYN output includes a
synthetic spectrum for the system, the WD, and the secondary star, all for an adopted orbital longitude and
orbital inclination, $i$.

\section{WX LMi}

WX LMi was found in the Hamburg Quasar Survey by Reimers et al. (1999)
who determined the orbital period, the spectral type of the secondary and
the distance. From the spectrum and photometry, they concluded there
were two accretion spots, with magnetic fields of 60 and 68 MG and
the accretion rate was about 100 times less than normal polars.
Further long term photometry was obtained by Schwarz et al. (2001)
and modeled to determine the colatitudes (90-130$^{\circ}$) and azimuths 
(-70 to -80, and +90$^{\circ}$) of the spots. These locations
placed the spots
slightly below the equator of the white dwarf and further from
the line-of-centers between the stars than in normal high accretion rate polars.
V07 used {\it XMM-Newton} to obtain X-ray data as well as some near UV
photometry using the optical monitor (OM) UV filters
at 2910 and 3440\AA. They also used optical spectra 
to establish an upper limit to the temperature for the 
white dwarf of 8000K and to constrain the cyclotron model and accretion
rate. They concluded that the low accretion rate was consistent with  
accretion resulting from 
wind loss from the secondary. Their model found small (40 km
radius) accretion spots at similar locations to that of Schwarz et al. (2001)
but the X-ray vs UV modulations and the hardness ratios were not consistent
with heated spots. The X-ray emission was assumed to originate from
both coronal activity on the secondary and from the accretion regions. 

The {\it GALEX} data explore the regime between the X-ray and OM
filters. The UV light curves (with 15 phase bins) 
are shown in Figure 1 along with the optical data
obtained about 2 months later. The double-humped light curve variation
that has been taken to be evidence of the accretion areas undergoes an
interesting variation with wavelength. Figure 1 shows the progression from
a double-humped curve in the I filter to a very low level of variability in
the optical $B$ and then an increasing variation as the FUV is reached.
The phasing here is spectroscopic (phase 0 is inferior conjunction of the
secondary) so the primary pole is evident between
phase 0.8 and 0.9, the secondary pole near phase 0.3 and both poles are self-eclipsed
near phase 0.6. The FUV shows slightly higher flux for the
secondary pole than the primary, which corroborates what V07
regarded as a puzzle i.e. that the secondary spot seems hotter than
the primary one. The dips at phase 0.3 in $V$ and $B$ are related to
the cyclotron features at 5950\AA\ and 5220\AA\ which appear in these filters
and change strength during the orbit due to the changing viewing angle.

\subsection{Model Parameters} 

V07 determined or adopted some of the parameters needed for a simulation 
of the WX LMi system including the orbital period, $P$; $M_{\rm wd}=0.6M_{\odot}$(adopted); 
$M_{\rm s}=0.179M_{\odot}$;
$T_{\rm eff,wd}=8000$K; and $T_{\rm eff,s}=3300$K, appropriate to an M4.5 spectral type. They also determined some of the spot parameters associated with
the magnetic field, including $\beta(1,2)=145, 135$, the spot colatitudes, and 
$\psi(1,2)= 60,-95$, the spot azimuths. These 
parameters are listed in Table 3.
The mass transfer rate ($\dot{M}$) in WX LMi ($\sim1.5{\times}10^{-13}M_{\odot}~{\rm yr}^{-1}$; V07), is listed as a system parameter in Table 3 but is
not needed in the analysis.

$M_{\rm s}=0.179M_{\odot}$, listed in V07 (Table~2) for a secondary star of spectral type M4.5,
differs slightly from the calibration of Knigge (2006; 2007), $M_{\rm s}=0.200M_{\odot}$ at the WX LMi
orbital period, but we preserve the value $0.179M_{\odot}$ for consistency with the other V07
parameters. We have assigned error bars in Table~3 from the range of $M_{\rm s}$ values in Table~2 of V07.
We adopt a secondary component model which fills its Roche lobe ($\Omega(\rm s)$, Table~3). 
V07 describe WX LMi as a pre-polar candidate, appropriate to the very small $\dot{M}$, and underfilling
its Roche lobe. It is
convenient for our analysis to assume a secondary star which fills its Roche lobe; after obtaining a
solution we tested the sensitivity of our assumption by performing a simulation in which the secondary
underfills its Roche lobe. 

The adopted value of $M_{\rm wd}=0.6M_{\odot}$ has no assigned error bars. The radius of a zero temperature 
homogeneous Hamada-Salpeter carbon $0.6M_{\odot}$ WD is $0.0120R_{\odot}$ 
(Panei et al. 2000); we adopt this 
value with no
change for the very small correction to a 8000K WD. The WD Roche potential, $\Omega_{\rm wd}$, Table~3, 
produces the required $R_{\rm wd}$.
The Table~3 $T_{\rm eff,wd}=8000$K was determined by V07 but they state that the true value
is less than 8000K; they were unable to pursue this topic because of their lack of lower temperature models.
 V07 show that $i$ is less than $72{\arcdeg}$; the presence of detectable ellipticity
in the light curve suggests a fairly large $i$; we adopt $i=70{\arcdeg}$. Figure~2
 shows the final system synthetic
spectrum at orbital phase 0.0. This synthetic spectrum  can be compared with the
observed spectrum (Figure~1 of V07).

\subsection{$B, V, R, I$ Synthetic Photometry light curves}

We calculated and stored system synthetic spectra for 33 orbital longitudes for each test simulation. Each 
synthetic spectrum extended
from 800\AA~to 10,000\AA~and included Eddington flux values at 1\AA~spacing, produced by interpolation within
the input spectra to BINSYN and integrated, within BINSYN, over the photospheric segments visible to the observer, 
and with an adopted limb darkening of 0.6. Our initial simulation neglected any hot spots on the WD as we expected  
possible hot spots only to make a
negligible contribution to the $B,V,R,I$ light curves. We subsequently verified this expectation.

The APO filter transmission curves and the quantum efficiency curve of the CCD
used for the ground-based photometry were digitized by hand. The CCD quantum efficiency drops to 0.0 at 10,000\AA~ and this sets
the upper wavelength limit for the simulation. We also produced comparable transmission curves for the
FUV and NUV filters using the {\it GALEX} calibrations\footnote{http://galexgi.gsfc.nasa.gov/docs/instrument.html}.

Program SYNPHOT, within the BINSYN package, produces the integrated physical flux for each filter resulting 
from a system
synthetic spectrum incident on the APO filter set and CCD, given by the expression 

\begin{displaymath}
F_s = \frac{\int S(\lambda)f(\lambda)d\lambda}{\int S(\lambda)d\lambda}
\end{displaymath}

where $S(\lambda)$ is the photometric system response function and $f(\lambda)$ is the synthetic
spectrum incident flux.

Numerical integration requires a value for the filter transmission and CCD response at each tabular
wavelength of the system synthetic spectrum. We used the Numerical Recipes (Press et al. 1986) 
program {\bf Spline} to fit each
digitized filter transmission curve as well as the CCD response curve, and the Numerical Recipes
routine {\bf Splint} to produce the required transmission or response at the synthetic spectrum wavelengths.

The quantity to compare with an observed light curve is the integrated flux divided by $D^2$,
where $D$ is the distance to the system in cgs units. We initially adopted the V07 determination of 100~pc.
The initial test showed rough agreement between simulation and observed light curves with variable
cyclotron contributions. The calculated light curves were systematically too bright in $B, R$, and $I$, with
a strong cyclotron contribution in $V$. We varied the assumed distance to obtain an optimum overall fit
by eye and determined a revised distance of 107~pc. This value is in very close agreement with
the V07 value of 104~pc in their Table~2 for the M4.5 spectral type adopted for the secondary star in this
study.

The final light curves, compared with the observations, are shown in Figure~3 
(note the change in ordinate scale in Figure~3d).
Cyclotron emission dominates the $V$ band light curve. This is in agreement with the V07 cyclotron model, their
Figure~7. The same Figure~7 indicates some cyclotron contribution to the $B, R$, and $I$ light curves. 

Figure~3d shows that the synthetic light curve has slightly too much flux for the adopted distance of 107~pc.
As Figure~2 shows, the secondary component spectrum rises rapidly to a peak near 10,000\AA. A small departure
of the actual WX LMi secondary component from the adopted model could explain this
residual. 

The inclination cannot be increased more than a few degrees without producing eclipses; a reduction in the
inclination decreases the light curve amplitude and increases the absolute flux level. We regard the adopted
$i=70\arcdeg$ as optimum, considering the light curve contributions by cyclotron emission.

\subsection{NUV and FUV Synthetic Photometry light curves}

Figure~7 of V07
makes no prediction concerning cyclotron emission in
the FUV or NUV, and the contribution of the secondary
component becomes negligible shortward of about 3500\AA\ so it
 does not contribute in the NUV or FUV. 

We adopted circular models for the two hot spots and experimented with their $T_{\rm eff}$ values and angular
radii. Initial simulations suggested a WD $T_{\rm eff}=7800$K. Initial choice of spot $T_{\rm eff}$
values of 12,000K produced a FUV light curve amplitude that was much too large. 
Lower temperature spots require larger radii to preserve the light curve amplitude, particularly
in the NUV.
Eventually we found an acceptable
fit with spot radii of $32\arcdeg$ and $T_{\rm eff}$ values of 10,000K. A final correction of the WD 
$T_{\rm eff}$ to 7900K produced the best fit, by eye (Figure~2 has a
WD with $T_{\rm eff}$ of 7900K).
The FUV and 
NUV light curves are shown in Figure~4 and 
the final spot parameters are in Table~4. 

The spots are large; a view of the WD at orbital phase 0.0 is in Figure~5. We suggest that this result 
is not inconsistent with the small $\dot{M}$ and impacting stream from the secondary component wind
rather than a stream emerging through the L1 point. While our spot sizes are
much larger than those of V07, (their estimate of the accretion area was about
40 km in radius, whereas our 32$^{\circ}$ spots are about 4700 km in radius),
their estimate was based on the X-ray and beamed cyclotron regions. These areas
are typically much smaller spots than the heated areas visible in the UV
(G\"ansicke et al. 2006). Time-resolved UV spectropolarimetry will be required to obtain further
 knowledge of cyclotron effects in the UV. For the field strength of 68 MG, the
 n=6-8 harmonics will be in the NUV band. This could account for some  of the
deviations of the fit in this band. 

 As a check, we recalculated the $B, V, R$, and $I$
light curves with and without the spots included in the model. There was a barely detectable spot
contribution to those light curves, far too small to require a revision of the model. Our model does
not prove that hot spots are the cause of the NUV and FUV light variation, it
only demonstrates that
our adopted parameters can approximate the observed light variations.
The
deviations likely indicate that the spots have more complicated geometries
 than our
simple assumption of a circular shape.

\subsection{Does the secondary component fill its Roche lobe?}

Our model assumes the secondary component fills its Roche lobe. As a test we changed the value of 
$\Omega_{\rm s}$ (Table~3), so that the volume of the secondary was 95\% of the 
initial model value which filled the Roche lobe.
Recalculation of the $B, V, R, I$ light curves showed that the amplitudes are
almost unaffected. A slight amplitude reduction of the $I$ light curve is barely
detectable. Further reducing the volume to 90\% produces the changes shown in
Figure 6. Thus, the secondary can underfill the Roche lobe by a small amount without
reducing the
theoretical light curve amplitudes sufficiently to be in disagreement with
the observed
light curve. However, if it underfills by a large
factor,
the light curve amplitude, especially in $I$, would be too small
due to the
reduced ellipticity of the secondary. 
 If there is some contribution of the n=2 harmonics in the $I$ filter, the
cyclotron would dilute the ellipsoidal variation from the secondary, which
would necessitate an even larger filling factor, although the variation from
the cyclotron would then add to the variablility as well.

\section{SDSS1031}

As with the majority of the LARPS, SDSS1031 was discovered by searches through the
SDSS spectroscopic survey (S07). The variations in the polarimetry and photometry
determined a short orbital period and the spacing of the cyclotron harmonics
revealed the magnetic field. The spectral fluxes between the
cyclotron harmonics placed limits on the white dwarf temperature and
the late spectral type of the secondary (Table 1 summarizes these values).
Its very short period makes it unusual among the LARPS (Figure 5 of Sw09),
and may present problems for its evolution as a PREP. However,
its low accretion rate, cool white dwarf and late type secondary
which underfills its Roche lobe (S07) all match the primary characteristics of
LARPS. SDSS1031 shares many characteristics of the short period 
normal polar EF Eri which
has mostly been is a low state since 1997 but does have occasional high
states (Howell et al. 2006).

The UV through optical light curves of SDSS1031 are shown in Figure 7, where the
arbitrary phase in UV and optical is set so that maximum light occurs
roughly at phase 0. While SDSS1031 is the faintest of our three sources
in the optical, the FUV magnitudes are comparable to WX LMi.
The $R$ filter shows the cyclotron variation found
by S07 with some evidence of a double-humped shape indicative of beaming in 
an optically thin high cyclotron harmonic. The $V$ filter shows the least orbital
variation while $B$ has an increased amplitude. The amplitude of the UV orbital 
modulation
is much lower than in WX LMi and more asymmetical in shape,
showing a gradual rise to maximum followed by a steeper decline. If the
variation is due to a hot spot, the
similar amplitudes in FUV and NUV mean that the spots must be different
sizes and simple circular isothermal spots will not provide a good fit
to the observations. To get a crude idea of what kind of spot would
be needed, the BINSYN model was computed for the parameters shown in
Table 4 and fit to the {\it GALEX} data binned by 10 in phase (Figure 8). The
best fit obtained was with a 13,000K spot at -65$^{\circ}$ latitude.
The spot radius was 10$^{\circ}$ for the FUV and 23$^{\circ}$ for NUV.
However, as expected, the shape of the variation is not fit well.

\subsection{SDSS1212}

Like SDSS1031, this object was found in searching through the SDSS spectra (Schmidt et al.
2005b) because the Balmer lines showed Zeeman splitting from a magnetic
white dwarf as well as weak H$\alpha$ emission. Spectroscopy by these authors
revealed a period near 90 min but no secondary star was detected 
and an IR measurement
at $J$ implied a brown dwarf object of spectral type L5 or later. Further
IR observations by Debes et al. (2006) refined the orbital period and
showed a variation in the $K$ band that could be explained  with cyclotron
emission from a small spot on the white dwarf. Farihi, Burleigh \& Hoard
(2008) modeled further IR data with an L8-T1 substellar dwarf and a white dwarf 
with a 7 MG field. B06 obtained X-ray
observations with {\it Swift} which detected the source but with too
low of a count rate to ascertain variability. The flux was best fit with
 a thermal plasma with kT=1.9 keV and a luminosity of 10$^{29}$ erg s$^{-1}$.
While this luminosity is comparable to that detected in the LARP MQ Dra
(Szkody et al. 2004), and the low state of the polar AR UMA (Szkody et al.
1999), it is far larger than the X-ray luminosities of L dwarfs (Stelzer et al.
2006). The $u,g,i$ photometry of B06 shows a single hump that increases
in amplitude in the blue, with some slight indication of a second small peak
half a phase later that was pointed out by Koen \& Maxted (2006). The origin of
the second peak is not clear; Koen \& Maxted (2006) suggested a reflection effect
from the secondary (since the feature seemed to increase at longer wavelengths), 
while B06 mention a possible second pole (but this is not
evident in the IR photometry of Debes et al. (2006).

B06 model the primary optical hump as a 14,000K hot spot on the white
dwarf. However, as they point out, it is not clear what provides the accretion
necessary to produce the X-rays and a hot spot. A brown dwarf that underfills
its Roche lobe should not produce an accretion stream nor a wind outflow.
Recent photometry by Howell et al. (2008) showed that the strength of H$\alpha$
is variable, indicating that the secondary has some activity. It may be that
the secondary stars in close binaries have higher activity levels than their
single counterparts due to their faster rotations in a binary system.

Our {\it GALEX} data on SDSS1212 (Figure 9) shows a similar single-peaked
modulation with increased amplitude in FUV over the NUV and a possible hint of
a second small peak near phase 0.5. The NUV magnitude
is the brightest of the three objects. Trial models with BINSYN
using the B06 parameters for the $i$, masses, and temperature of
the white dwarf revealed that the width of the hump is a clear function of
the spot latitude, with greater width when the spot is in the same hemisphere
as the visible pole. A good fit (Figure 10 bottom) to the NUV light curve was
obtained with a 14,000K spot (the same temperature found by B06 from the
optical light curves) with a radius of 9$^{\circ}$, but this spot
produced a FUV amplitude that was too large. Thus, as in MQ Dra (Szkody
et al. 2008), a single spot size and temperature cannot fit both the NUV
and FUV. If the spot temperature is kept at 14,000K, a smaller spot
of 7$^{\circ}$ fits the FUV reasonably well (Figure 10 top). The observed
light curve shows some scatter and there is a hint that the light curve
is asymmetric with a slower rise than decline. The spot sizes are
1\% of the white dwarf surface in FUV and 1.7\% in NUV, as compared to the
5\% found by B06 from their optical modeling. These numbers follow the trend of
decreasing spot sizes with smaller wavelengths.

Increased time-resolution and S/N will be needed to further understand the
reality and nature of the possible peak near phase 0.5. It is intriguing that
this feature is evident from UV through the $R$ band. These phases need to be
targeted to determine if there is accretion related to a second pole.

\section{Conclusions}

Our UV observations of three polars with extremely low accretion rates
enforce the results found from observations of EF Eri and MQ Dra i.e.
that all the white dwarfs have areas of enhanced emission even with
these low rates, indicating some accretion is still occuring. Despite a large range
in magnetic field (7-70MG), the field is apparently strong enough in 
all three systems to funnel the accreting material to the magnetic pole(s)
of the white dwarf. This appears to happen even in the case of SDSS1212,
which likely has a brown dwarf secondary. Thus, if the accretion is via a stellar
wind from the secondary that is trapped by the field of the white dwarf, it is
difficult to provide a wind of this type for SDSS1212. Time-resolved
spectra that would enable Doppler tomography might provide a resolution of
this issue. During high states of accretion, the stream flow is usually
visible as a narrow component in the Balmer emission lines which is mapped
to the stream. In normal polars during their low states, high velocity components
in the lines have been interpreted as structures similar to prominences close
to the secondary star (Kafka et al. 2007, 2008). However, the Balmer emission 
in LARPS is weak to non-existent due to the low
the accretion rates low so it would be difficult to construct
the map.

If the increased emission evident at some phases is
due to hotter temperature, simple spots of 10-000-14,000K covering a
few percent of the white dwarf surface can approximate the UV light curves,
although the geometries of the spots require a more complex model shape
than simply circular. Alternatively, if there is a large range in field strengths
in the white dwarfs, there could be contributions to the UV from higher
cyclotron harmonics. Improved UV cyclotron models will be needed to test
this possibility. Since these objects are too faint for {\it GALEX}
spectra or high S/N time-resolved UV light curves that would merit detailed
models of the shape or of pursuing cyclotron radiation, further work
will require UV spectra and polarimetry with larger telescopes.

\acknowledgments

Support for this research was provided by NASA {\it GALEX} grant NNX08AM07G.
We gratefully acknowledge Pierre Bergeron for providing cool white dwarf
models, the MARCS database for access to their spectra, Eric Bullock for work 
with some of the data files, and
Mark Klaene for providing optical filter curves.

\clearpage

\begin{deluxetable}{lccc}
\tablewidth{0pt}
\tablecaption{Objects}
\tablehead{
\colhead{Parameter} & \colhead{WX LMi} & \colhead{SDSS1031} & \colhead{SDSS1212}}
\startdata
P$_{orb}$ (hr) &  2.78 & 1.37 & 1.47 \\
Optical Mag & $V$=16.97 & $g$=18.26 & $g$=17.99\\
T$_{wd}$ (K)  & $<$8000 & 9000 & 9500 \\
B (MG) & 61,70 & 42 & 7 \\
Sec Type & dM4.5 & $\ge$dM6 & L8-T1 \\
d (pc) & 100 & 270-380 & 120 \\ 
{$\dot{M}$} ($\times$10$^{-13}$M$_{\odot}$ yr$^{-1}$) & 1.5 & 1.5-3 & 1 \\  
Ref\tablenotemark{a} & 1, 2 & 3 & 4, 5, 6 \\
\enddata
\tablenotetext{a}{(1) Reimers et al. (1999), (2) Vogel et al. (2007), (3) Schmidt
et al. (2007), (4) Schmidt et al. (2005b), (5) Burleigh et al. (2006),
 (6) Farihi et al. (2008)}
\end{deluxetable}
 
\clearpage
\begin{deluxetable}{lccl}
\tablewidth{0pt}
\tablecaption{Summary of 2008 Observations}
\tablehead{
\colhead{Object} & \colhead{Date} & \colhead{UT} & \colhead{Data}}
\startdata
WX LMi & Feb 16 & 10:23-20:42 & {\it GALEX} NUV, FUV 7 visits\\
WX LMi & Apr 29 & 03:57-08:56 & $BVRI$ 39 exposures\\
SDSS1031 & Feb 23 & 10:01-20:19 & {\it GALEX} NUV, FUV 7 visits\\
SDSS1031 & May 5 & 05:04-08:15 & $BVR$ 12 exposures\\ 
SDSS1212 & Apr 5 & 11:14-23:01 & {\it GALEX} NUV, FUV 8 visits\\
\enddata
\end{deluxetable}

\clearpage
\begin{deluxetable}{llll}
\tablewidth{0pt}
\tablecaption{WX LMi Model System Parameters}
\tablehead{
\colhead{parameter} & \colhead{value} & \colhead{parameter} & \colhead{value}}
\startdata
${ M}_{\rm wd}$  &  $0.60{M}_{\odot}$    & $T_{\rm eff,s}$(pole)    &  $3300{\pm}100$K \\
${M}_{\rm s}$  &  $0.179{\pm}0.4{M}_{\odot}$     & $T_{\rm eff,s}$(point)       &       1640K\\
${\dot{M}}$      &  $1.5{\times}10^{-13}{M}_{\odot} {\rm yr}^{-1}$ &     $T_{\rm eff,s}$(side)  &       3259K\\
P    &  0.11592364 days   &      $T_{\rm eff,s}$(back)  &       3133K \\
$D$              &  $0.920131R_{\odot}$ & $r_s$(pole) &  $0.234R_{\odot}$\\      
${\Omega}_{\rm wd}$         & 76.65 & $r_s$(point)   & $0.348R_{\odot}$\\
${\Omega}_s$                &  2.4625438 & $r_s$(side)  & $0.250R_{\odot}$\\
{\it i}              &   $70\arcdeg$& $r_s$back)   & $0.280R_{\odot}$\\
$T_{\rm eff,wd}$         &  $7900{\pm}100$K  & log $g_s$(pole) & 4.95\\
$r_{\rm wd}$      &   $0.0120R_{\odot}$ & log $g_s$(point) & 1.16\\
log $g_{\rm wd}$  &   8.05& log $g_s$(side)  & 4.89\\
$A_{\rm wd}$        &  1.0 & log $g_s$(back)  & 4.67\\ 
$A_s$               &  0.6 & ${\beta}$(1,2) & $145, 135\arcdeg$  \\
${\beta}_{\rm wd}$  &  0.25 & ${\psi}$(1,2) & $60, -95\arcdeg$ \\
${\beta}_s$         &  0.08\\
\enddata
\tablecomments{${\rm wd}$ refers to the WD; $s$ refers to the secondary star.
$D$ is the component separation of centers,
${\Omega}$ is a Roche potential. 
$A$ values are bolometric albedos,  ${\beta}_{\rm wd}$ and ${\beta}_s$ values are 
gravity-darkening exponents, ${\beta}$(1,2) and ${\psi}$(1,2) are spot colatitudes
and spot azimuths. See the text for a discussion.
} 
\end{deluxetable}

\clearpage
\begin{deluxetable}{lccc}
\tablewidth{0pt}
\tablecaption{Model Parameters}
\tablehead{
\colhead{Parameter} & \colhead{WX LMi} & \colhead{SDSS1031} & \colhead{SDSS1212}}
\startdata
WD T$_{eff}$  & 7900K & 9500K & 9500 \\
WD mass (M$_{\odot}$) & 0.6  & 0.6 & 0.6 \\
WD radius (R$_{\odot}$) & 0.012 & 0.012 & 0.012 \\
WD log g &  8.0 & 8.0 & 8.0\\
Sec mass (M$_{\odot}$) & 0.18 & 0.08 & 0.08 \\
i ($^{\circ}$) & 70 & 70 & 70 \\
Spot T$_{eff}$  & 10,000; 10,000K & 13,000K & 14,000K \\
Spot latitude & -55; -45$^{\circ}$ & -65$^{\circ}$ & -15$^{\circ}$ \\
Spot ang rad FUV & 32$^{\circ}$; 32$^{\circ}$ & 10$^{\circ}$ & 7$^{\circ}$ \\
Spot ang rad NUV & 32$^{\circ}$; 32$^{\circ}$ & 23$^{\circ}$ & 9$^{\circ}$ \\
\enddata
\end{deluxetable}

\clearpage
\begin{figure} [h]
\figurenum {1}
\plotone{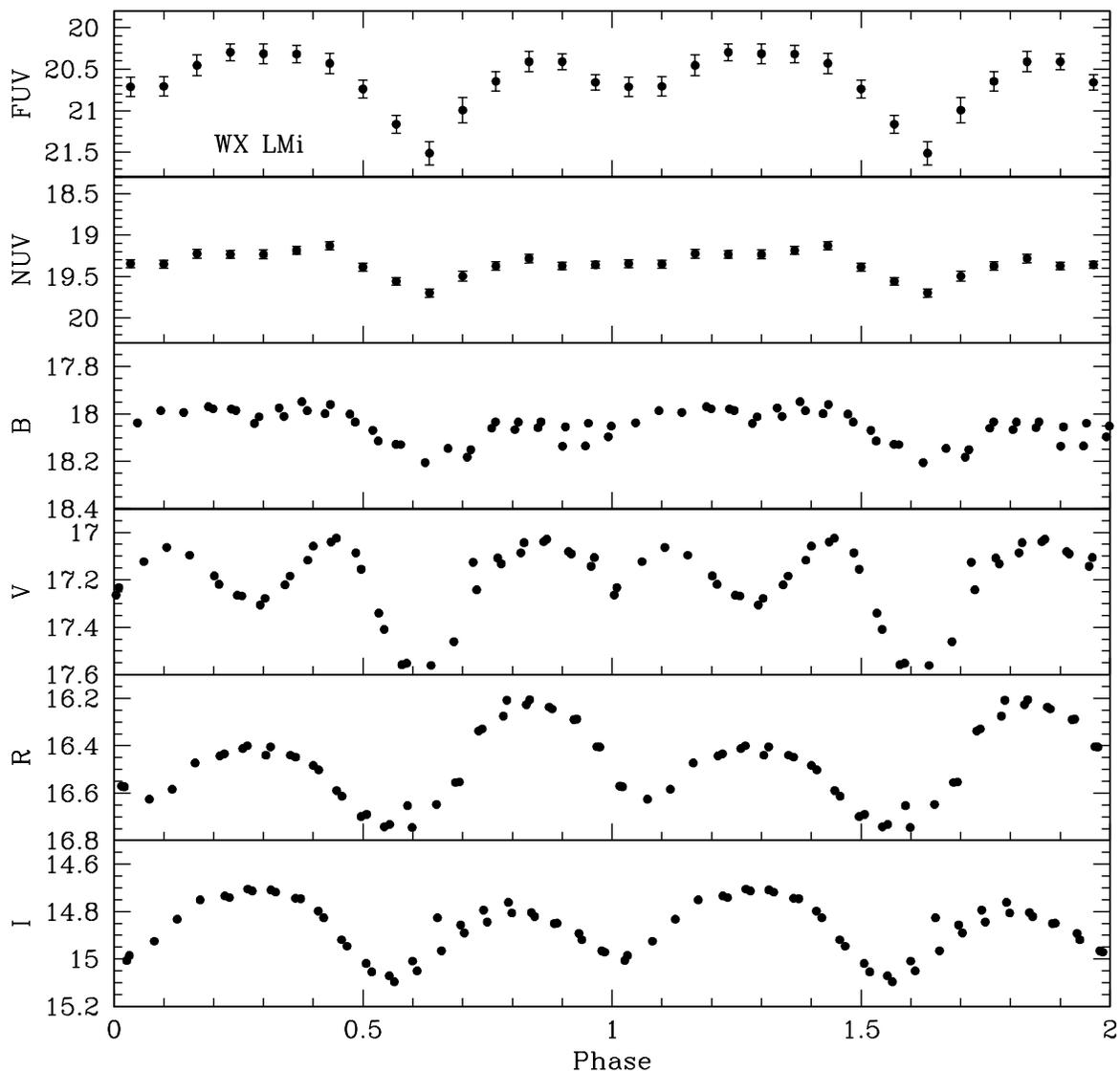}
\caption{{$\it GALEX$} FUV and NUV and optical $B,V,R,I$ filter light curves as a 
function of phase (phase 0 is inferior conjunction of 
secondary from V07) for WX LMi. Light curves are repeated from phases 1.0 to 2.0.
Error bars for the optical data are smaller
than the points. The comparison star for $B, V, R, I$ is star A from Schwarz et al. 2001.}
\end{figure}

\clearpage
\begin{figure}
\figurenum {2}
\plotone{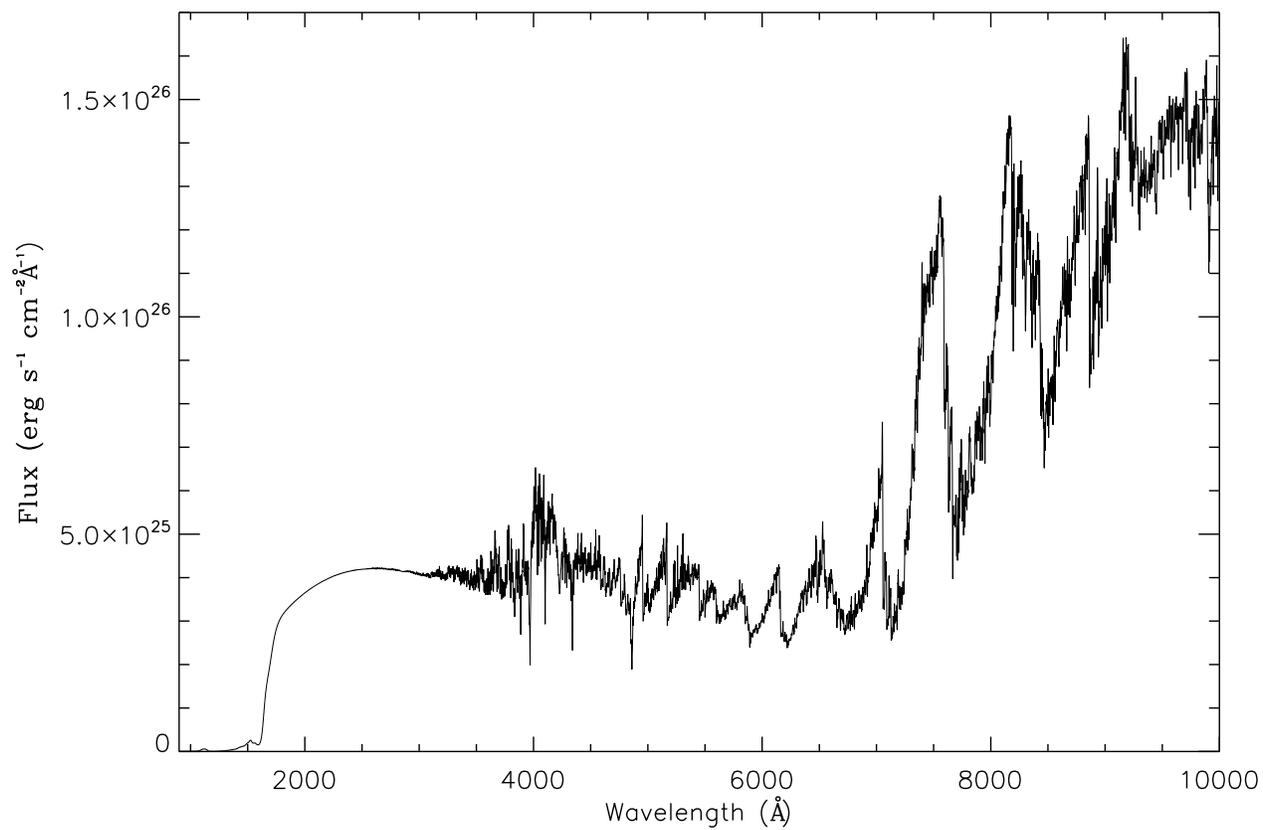}
\caption{Synthetic spectrum of WX LMi at orbital phase 0.0. The flux is as 
seen at the star.}
\end{figure}

\clearpage
\begin{figure}[h]
\figurenum {3}
\epsscale{0.97}
\plotone{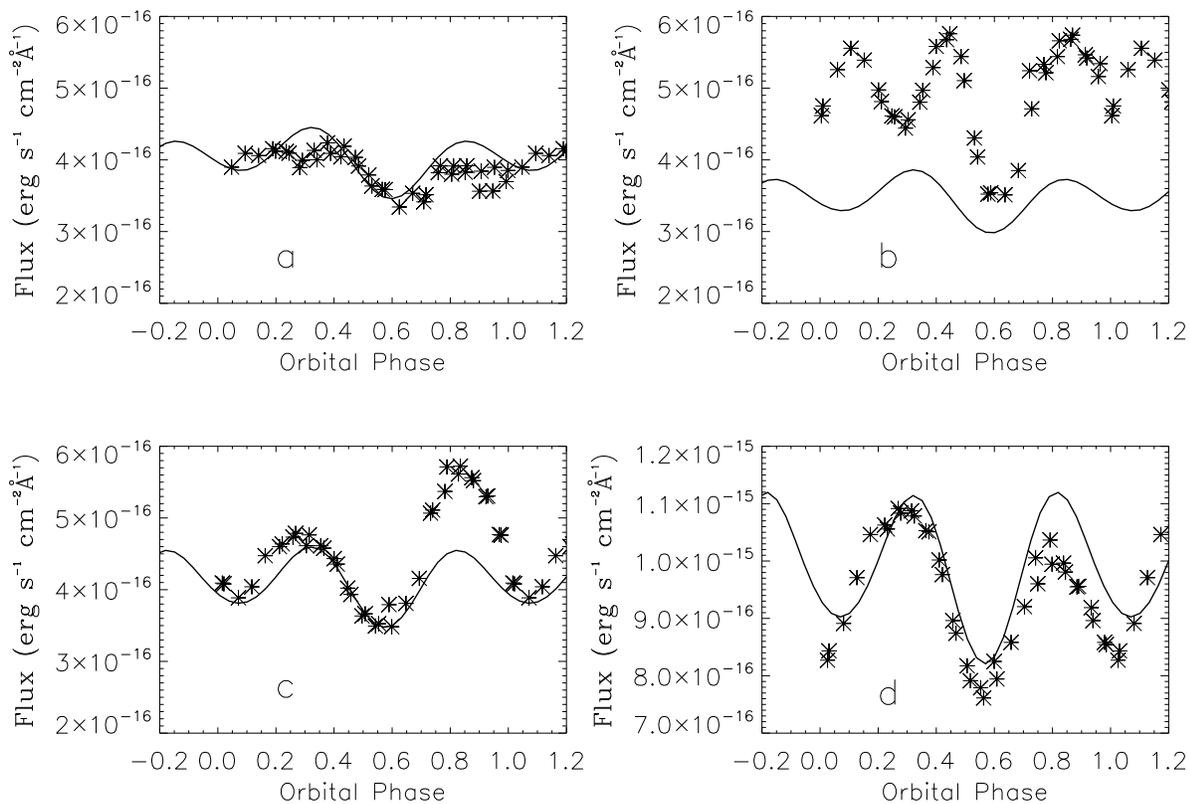}
\epsscale{1.00}
\vspace{5pt}
\figcaption{Synthetic photometry light curves of WX LMi as seen
from 107~pc; $B$ filter (a); $V$ filter (b); $R$ filter (c) and 
$I$ filter (d). Note the change in ordinate scale for the $I$ light curve.}
\end{figure}

\clearpage
\begin{figure}[h]
\figurenum {4}
\epsscale{0.8}
\plotone{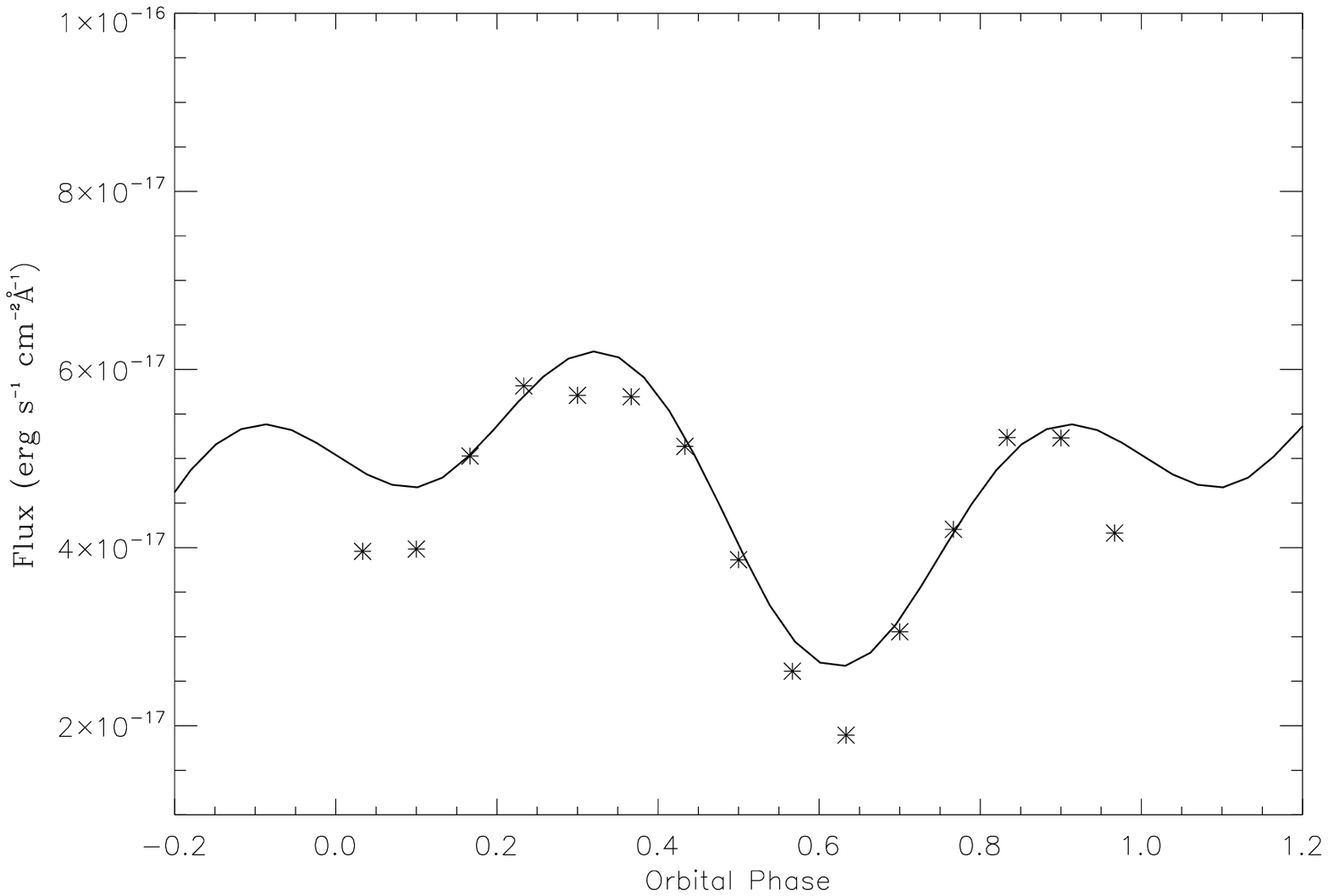}
\plotone{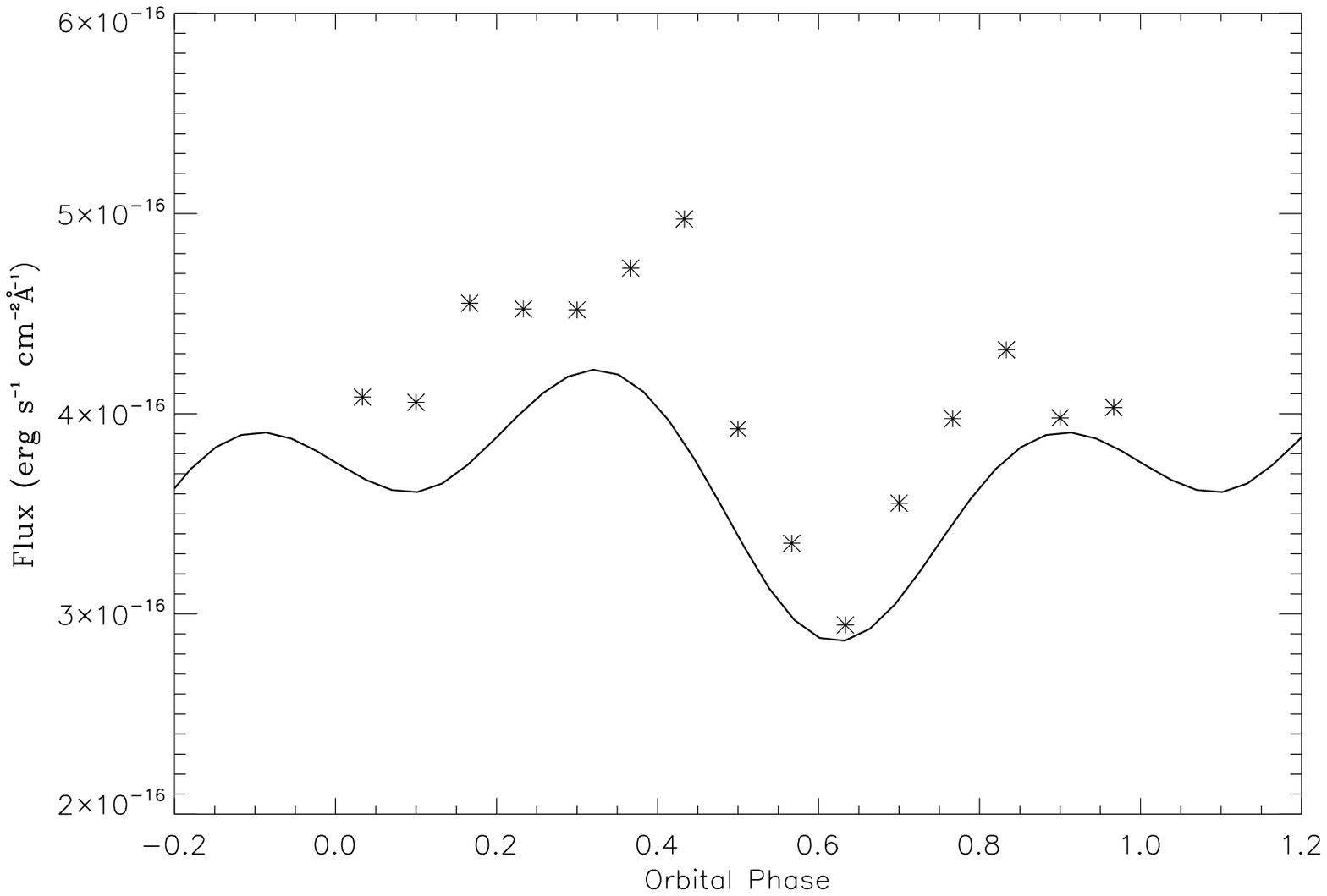}
\figcaption{Synthetic photometry FUV (top) and NUV (bottom) light curves
of WX LMi as seen from 107~pc. Note the different ordinate scales.}
\end{figure}

\clearpage
\begin{figure}[h]
\figurenum {5}
\epsscale{0.97}
\plotone{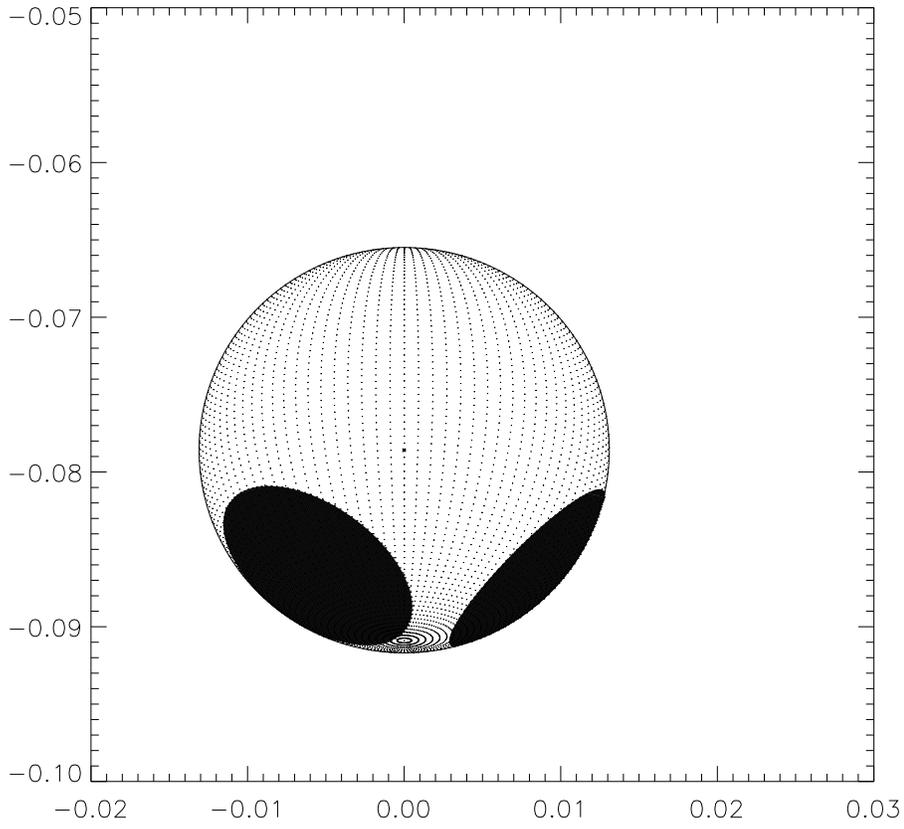}
\epsscale{1.00}
\vspace{5pt}
\figcaption{View of WX LMi white dwarf at orbital phase 0.0, showing
the two photospheric hot spots. See text for discussion.}
\end{figure}

\clearpage
\begin{figure}[h]
\figurenum {6}
\epsscale{0.97}
\plotone{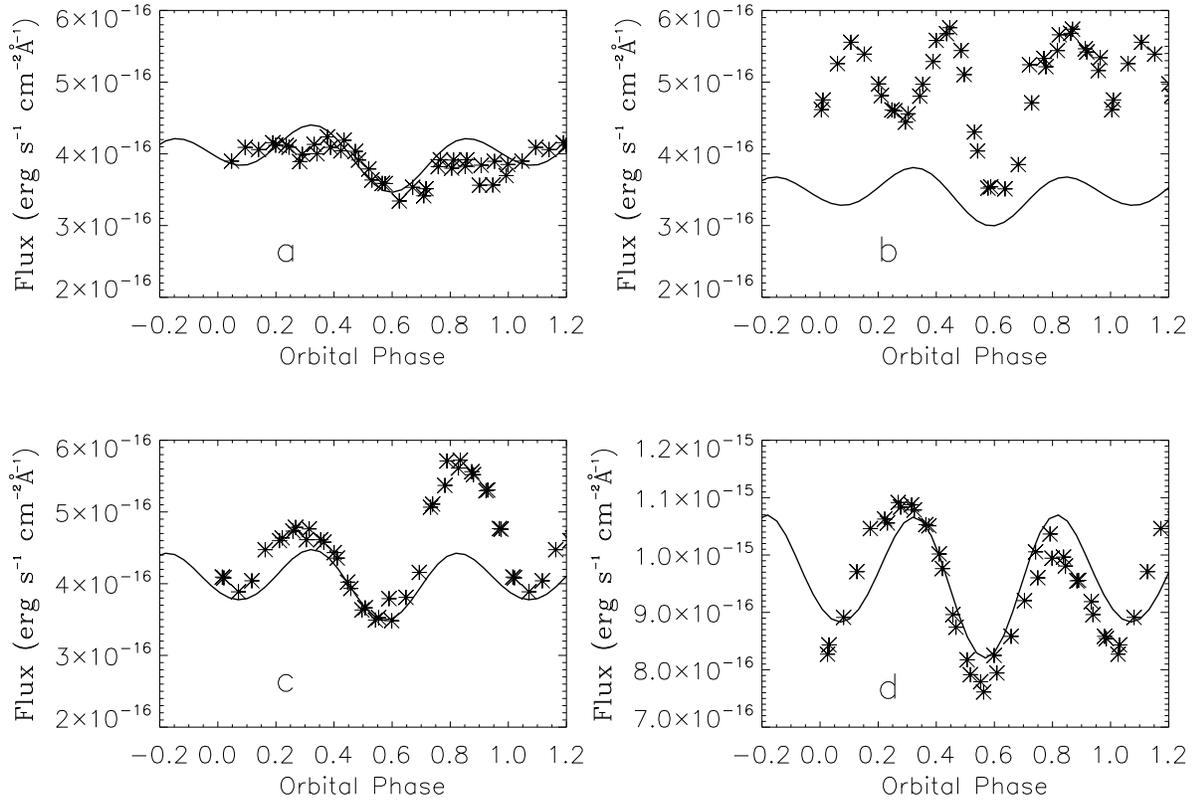}
\epsscale{1.00}
\vspace{5pt}
\figcaption{Synthetic light curves of WX LMi calculated with a secondary
underfilling its volume by 90\%; compare to Figure 3.}
\end{figure}

\clearpage
\begin{figure} [h]
\figurenum {7}
\plotone{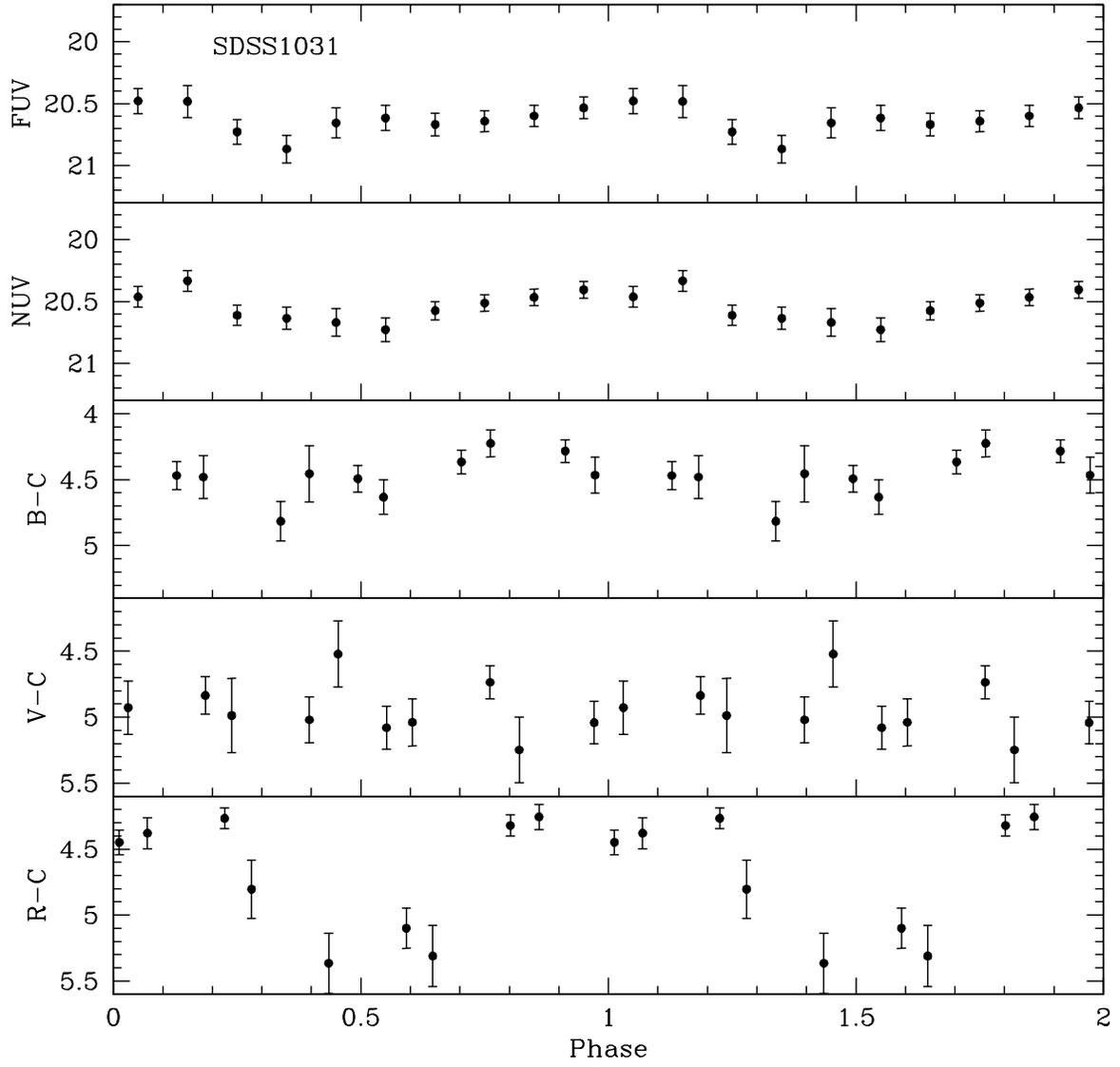}
\caption{{$\it GALEX$} FUV and NUV and optical $B,V,R$ filter light curves as a
function of phase (phases are arbitrary) for SDSS1031. 
Light curves are repeated from phases 1.0 to 2.0.}
\end{figure}

\clearpage
\begin{figure}
\figurenum {8}
\epsscale{0.8}
\plotone{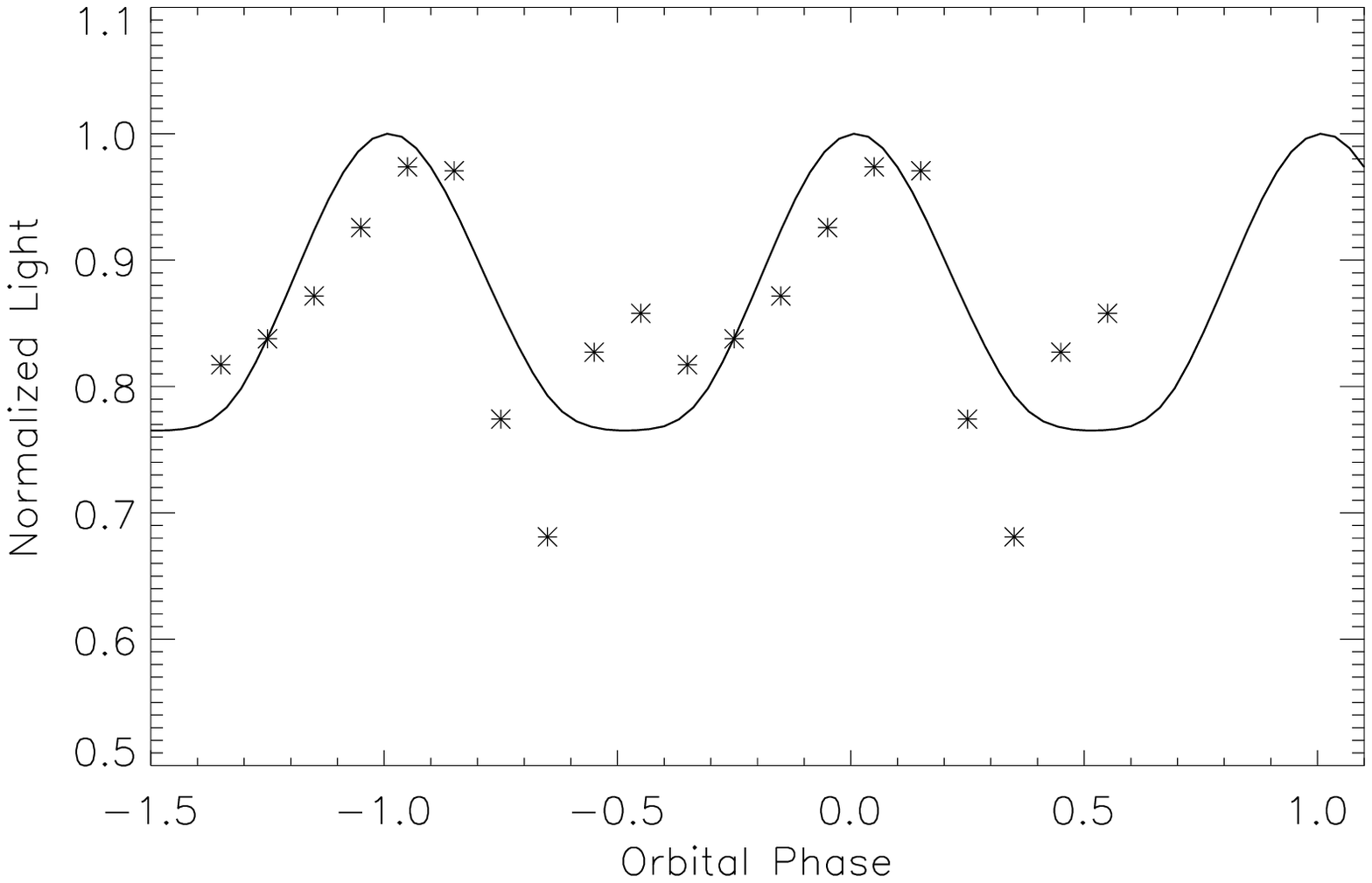}
\plotone{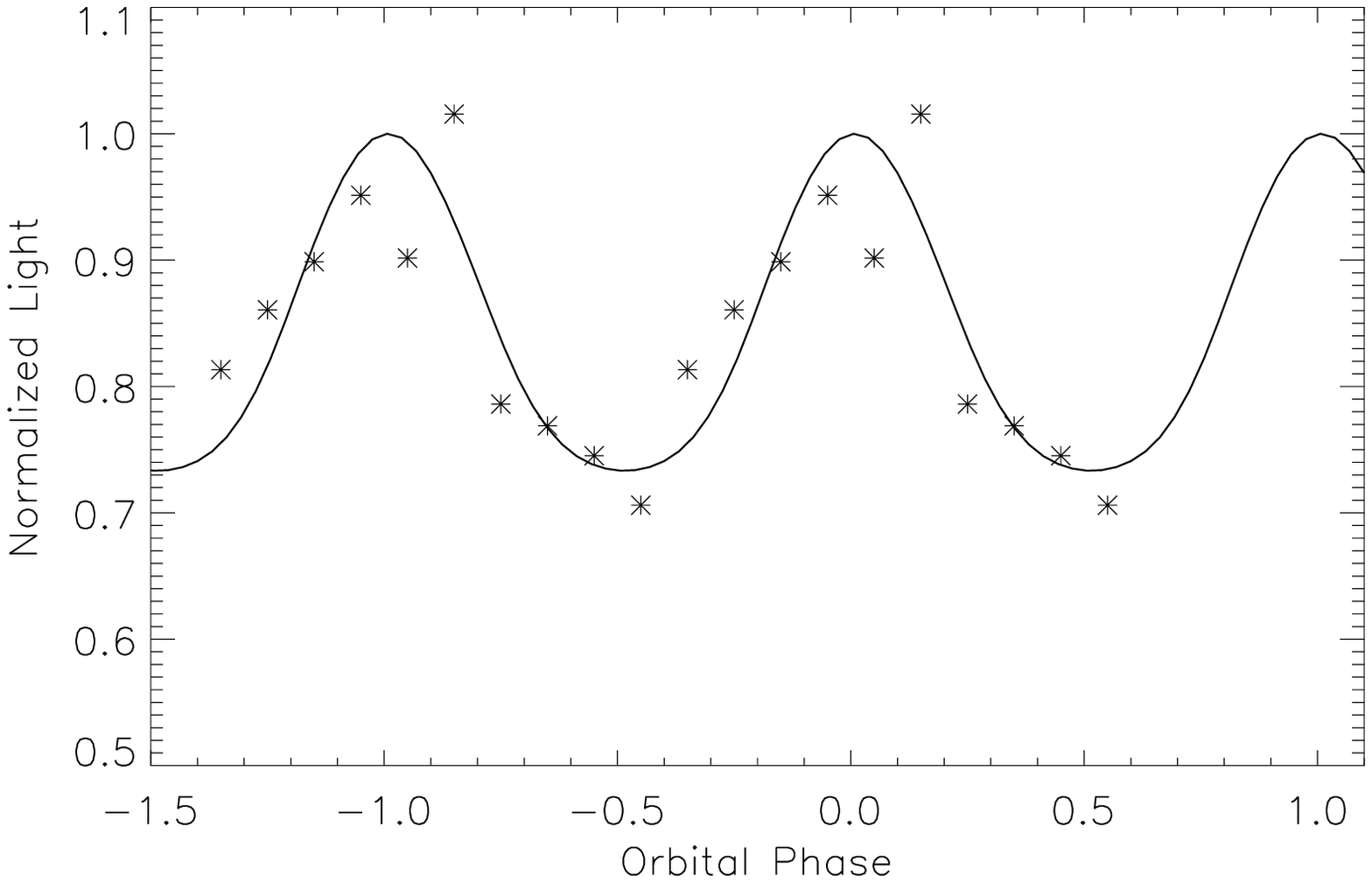}
\caption{Spot Model fits from Table 4 for the {$\it GALEX$} FUV (top) and NUV (bottom)
light curves of SDSS1031.}
\end{figure}

\clearpage
\begin{figure} [h]
\figurenum {9}
\epsscale{1.1}
\plotone{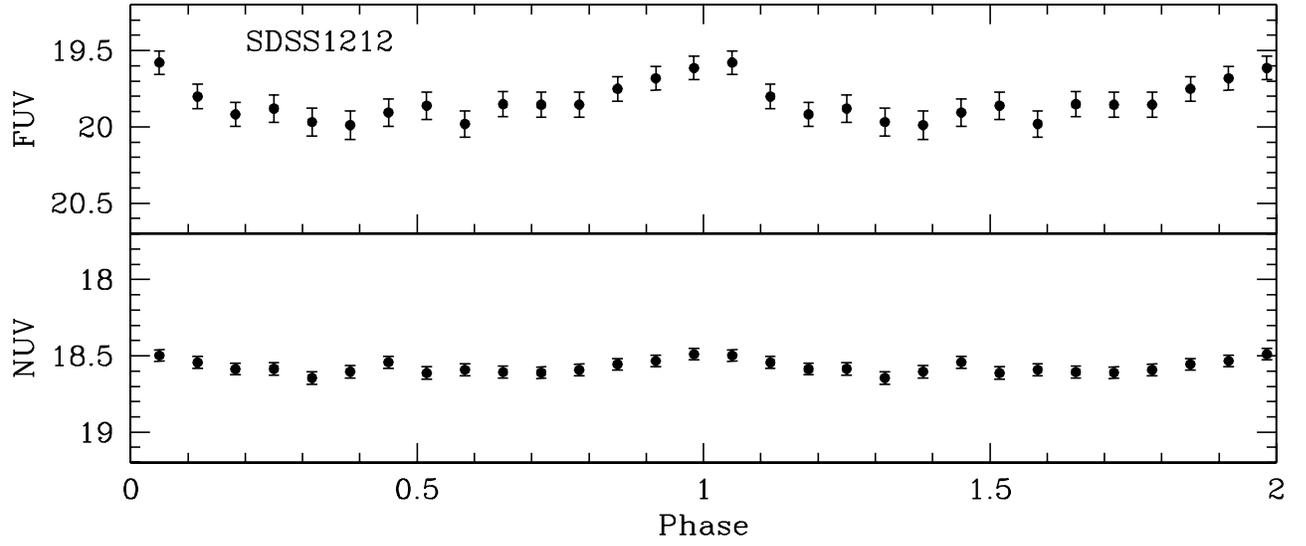}
\caption{{$\it GALEX$} FUV and NUV light curves as a
function of phase (phase 0 is optical photometric maximum with phasing
from B06) for SDSS1212.}
\end{figure}

\clearpage
\begin{figure}
\figurenum {10}
\epsscale{0.8}
\plotone{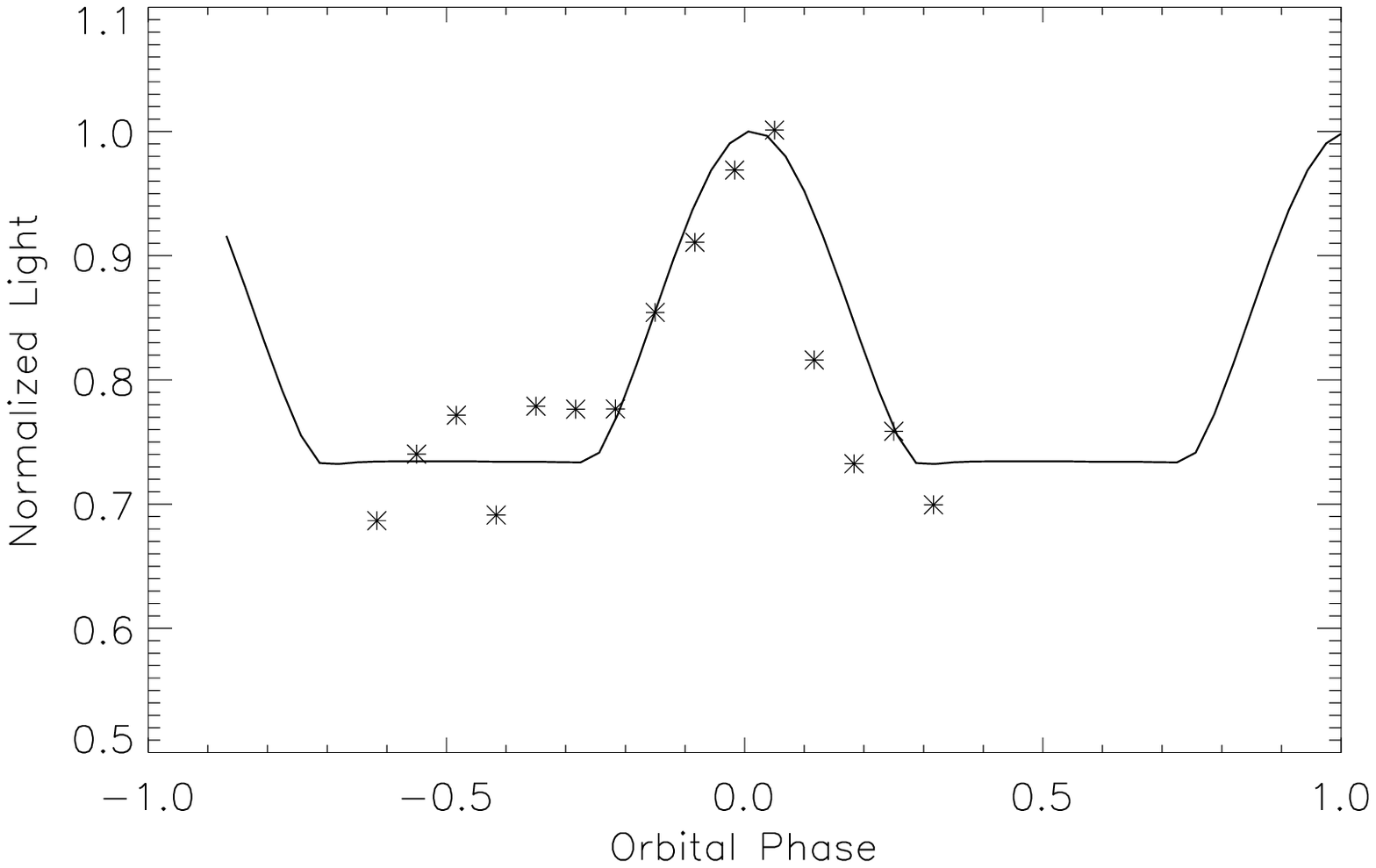}
\plotone{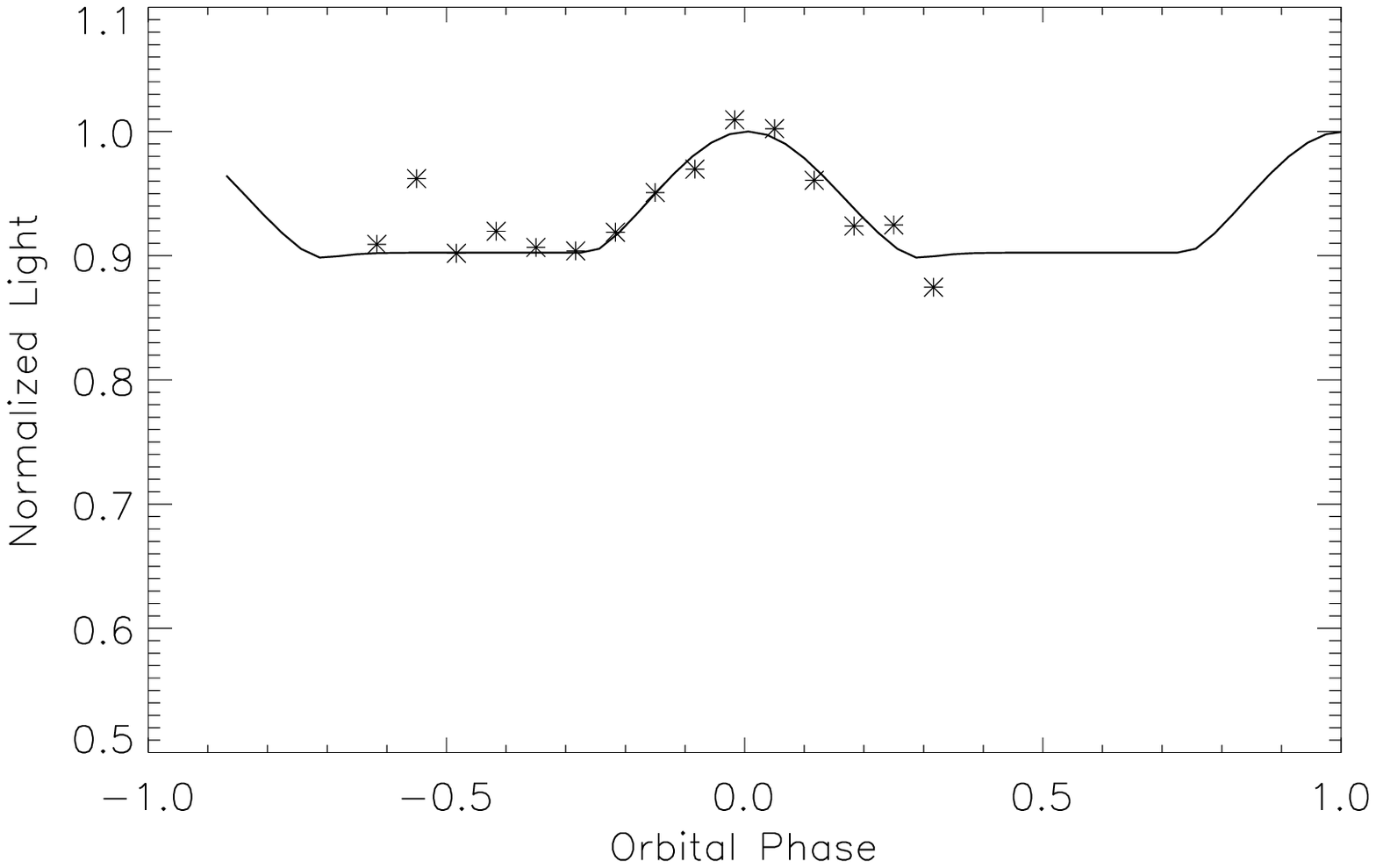}
\caption{Spot Model fits from Table 4 for the {$\it GALEX$} FUV (top) and NUV (bottom)
light curves of SDSS1212.}
\end{figure}

\end{document}